\documentclass{cimento}

\usepackage{graphicx,caption}  
\usepackage{wrapfig}
\usepackage{cite}
\usepackage{multicol}
\usepackage{amssymb}
\usepackage{url}
\title{Constraints on the density dependence of the symmetry energy}
\author{J.~\L{}ukasik\thanks{jerzy.lukasik@ifj.edu.pl}
for the ASY-EOS and ASY-EOS II Collaborations
}
\instlist{\inst{} IFJ PAN, PL-31342 Krak\'{o}w, Poland}

\begin{document}

\maketitle

\begin{abstract}
Current status and future experimental plans for constraining the symmetry
energy at supra-normal densities are presented. A special emphasis is put on
significance of the results obtained by the ASY-EOS Collaboration in a broader
astrophysical context, including the recent interpretations of the first LIGO
and Virgo GW170817 gravitational wave signal. The plans for a high
energy campaign at FAIR using the NeuLAND and the KRAB detectors will be
outlined.
\end{abstract}

\section{Introduction}

A lot of experimental and theoretical efforts have been made within more than a
decade now to study the properties of the asymmetric nuclear matter. Despite the
efforts, the underlying entity, the Equation of State (EoS) of nuclear matter
remains still uncertain, especially beyond the saturation density. In general,
the EoS relates the strength of the nucleon binding to the temperature, baryon
density, $\rho$, and isospin asymmetry $\delta$=($\rho_{n}$-$\rho_{p}$)/$\rho$
\cite{Dan02, Dan09}, where the subscripts $n$ and $p$ refer to the neutrons and
protons, respectively. For a cold matter the EoS is usually
split into a density dependent symmetric matter contribution and a symmetry
energy term proportional to the square of the asymmetry \cite{Lat00, Lat01,
LI08}: 
\begin{equation}
E(\rho,\delta)=E(\rho)+E_{sym}(\rho) \, \delta^{2}.  \label{eq_0} 
\end{equation}

\noindent The latter describes solely the dependence of the EoS on asymmetry and
is of high importance for both, the nuclear physics and the astrophysics.
Possible higher order terms with even powers of $\delta$ have been neglected
here. The problem has been attacked experimentally from three sides: by
performing the astrophysical and astronomical observations, by measuring the
observables related to the nuclear structure, and by investigating the density
and asymmetry dependent processes in heavy ion collisions. The nuclear and
astrophysical sources are expected to provide coherent results despite their
almost 19 orders of magnitude difference in scale. After all, they are driven by
the same nuclear interactions. But, how is it in reality?

Significance of the quest for the EoS is reflected by the amount and the scales
of the on-going and planned projects. The quest became not only 
multi-disciplinary but also a multi-messenger one, especially on the astronomy
and astrophysics side. Here, the astrophysical objects and processes are
observed using photons from radio waves through X-rays to gamma rays, neutrinos,
cosmic rays and gravitational waves. It is only recently that the ultra high
precision interferometers started to register the gravitational wrinkles due to
the neutron star, NS, merger events \cite{gw} and that the LISA Pathfinder
\cite{lisa} mission demonstrated that a space-based observatory of gravitational
waves is within our technical reach and can be operational around 2030. It is
also only last year that the NICER \cite{nicer} X-ray space observatory started
its mission at the ISS. The mission aims at accurate mass and radius
measurements of several NS through precise time resolved X-ray spectroscopy.
Such data should allow to precisely pick the corresponding EoS and pin down the
associated symmetry energy. The mission is supposed to operate for 18 months.
Taking into account that the predicted most probable rates for detection of
binary NS mergers with the advanced LIGO detectors are 10-500 events per year
\cite{rate}, one can expect a rapid progress in constraining the high density
EoS from the astrophysical sources in the nearest future.

The terrestrial laboratory investigations of the high density asymmetric nuclear
EoS become multi-messenger as well. The latest high energy experiment, carried
out in 2016 at RIKEN by the SPiRIT collaboration \cite{JHA16, LAS17}, aims at
extracting the symmetry energy by combining the information on charged pion
production rates, proton and neutron elliptic flows and possibly also on light
isobar flows. Some very preliminary results have already been presented this
year \cite{ISO18,JHA18,Miz18}.
The results of the two earlier high energy experiments: the FOPI-LAND
measurement \cite{lei93} reanalyzed by Russotto \etal \cite{rus11} and the
ASY-EOS measurement \cite{rus16} will be discussed below in a broader context.
Awaiting the beams from the upcoming FAIR facility a proposal by the ASY-EOS II
Collaboration of a new generation of an experiment will be presented.

\section{Symmetry energy}

The symmetry energy, $E_{sym}$, accounts for an excess of the nucleon binding in
a pure neutron matter with respect to a symmetric matter (the one with equal
numbers of neutrons and protons) at the same density. It is usually expressed in
the form of a Taylor expansion (\ref{eq_1}) around the normal density $\rho_{o}
\simeq$ 0.16 fm$^{-3}$:
\begin{equation}
E_{sym}(\rho) \simeq E_{sym}(\rho_{o}) + \frac{L}{3} \left ( \frac{\rho - \rho_{o}}{\rho_o}
\right ) + \frac{K_{sym}}{18} \left ( \frac{\rho - \rho_{o}}{\rho_o} \right )^2
 + \dots
\label{eq_1}
\end{equation}

\noindent where $L$ and $K_{sym}$ are the slope and curvature parameters at
$\rho_{o}$ and together with the value of the symmetry energy at $\rho_{o}$,
the  $E_{sym}(\rho_{o})$, they form a set of the main unknowns of the symmetry
energy. Apart from the higher order parameters there are, however, also other
quantities that enter a more exact parametrization of the $E_{sym}$ at the
mean field level when its momentum dependence is taken into account
\cite{bali17}. They appear in the form of effective masses which, in principle,
can be different for neutrons and protons. The $E_{sym}$ is usually also
split into its kinetic and potential parts, where the former is expressed using
the isoscalar effective mass which may be an additional source of uncertainty.

\section{Why so important?}

The density dependence of $E_{sym}$  is an important ingredient for evaluating
the drip lines, masses, density distributions and collective excitations of
neutron-rich nuclei in nuclear structure studies \cite{BRO00, ROC11}, flows,
fragment and particle production rates and multi-fragmentation in heavy-ion
collisions \cite{LI08,TSA17}, and also for simulations of astrophysical
processes like supernovae, stellar nucleosynthesis and objects like neutron
stars \cite{STE05}. 

In astrophysics the EoS in the density range 1-3 $\rho_{o}$ plays an essential
role in modeling the interiors of the NS \cite{Lat01}. It is still poorly known
in this range of densities but uniquely determines the  the relation between
their mass and radius, proton fraction, moment of inertia, crust-core
transition. Matter that is less susceptible to compression (described by a so
called stiff EoS) will favor larger NS for a given mass. Such an EoS also
predicts larger values for the maximum mass of a NS that can stand the
gravitational collapse into a black hole. On the other hand, easily compressible
matter (corresponding to a soft or super-soft EoS) will allow for smaller radii
and smaller threshold masses. As will be shown later, the current constraints on
the $E_{sym}$ allow for a still rather broad variation of the NS radii: from
$\sim$10 to $\sim$14 km.

\section{Why so uncertain?}

On the theory side the uncertainties arise, among others, from the fact that the
parameters of the phenomenological forces are being fixed around the saturation
density and for nearly symmetric matter while the extrapolations above
$\rho_{o}$ and for neutron rich or pure neutron matter have still a broad range
of freedom. At high densities the many body interactions begin to play a role
and here the uncertainties regarding their strength and isospin dependence
become noticeable. These and other deficiencies \cite{FUC06,GAN12, GAN16, BALI17a}
result in a broad spectrum of predictions for the nuclear EoS, even of those
resulting from the {\em ab initio} calculations, which are claimed to be
parameter free. Especially the $E_{sym}$ shows very different behaviors, in
particular at supra-normal densities, calling for more tight experimental
constraints at high densities. 

But, on the experimental side the situation is even more dramatic, because no
matter how precise the measurement is, extraction of the $E_{sym}$
parameters proceeds through some model inference and thus is model dependent.
This kind of circular dependence might look frustrating, but in fact it is not,
provided every other detail of the model, apart from the $E_{sym}$, is
well under control. Thus constraining the symmetry energy becomes a very
demanding task, requiring very high precision measurements and the state of the
art models. A very promising feature of the symmetry energy quest is that it is
multi-disciplinary and becomes also a multi-messenger one. This guarantees a
plenitude of complementary data from independent sources, which should overlap
and finally converge.

In reality, the way towards convergence is not at all simple. Recent
interpretations of the heavy ion data on the pion production rates can be an
example. Here, the FOPI data \cite{Rei07} on $\pi^{-}/\pi^{+}$ ratios have been
relatively well described by three different models \cite{Xia09}, \cite{Fen10}
and \cite{Xie13}, leading however to quite incoherent conclusions as far as the
stiffness of the $E_{sym}$ was concerned. This confirms the importance of
the efforts made by the transport model developers within the code comparison
project \cite{XU16} to perform thorough tests of the codes to identify their
weak points and understand the discrepancies.
Extraction of the EoS parameters from the astrophysical observations is not at
all easier. Fortunately, the new precise data provide new constraints.

\section{Mass and radius of neutron stars from a ``nuclear diesel''}

From a simple consideration of a stellar object in a hydrostatic equilibrium one
gets a dependence of the gradients of the pressure and of the mass shell on the
density, thus two equations and three unknowns, all depending on the radius. In
order to solve the problem one needs a third equation, the EoS, which relates
these quantities. Thus the EoS of neutron rich matter becomes indispensable for
realistic NS simulations. Since there is a unique correspondence between the
mass-radius relation and the EoS, it seems that measuring the masses and radii
of NS should be sufficient to pick the right one. While precise mass
measurements are possible especially for the radio and X-ray binary pulsars
\cite{Lat05,Dem10}, the measurements of the radii are more difficult and less
accurate. Most promising for simultaneous mass and radius measurements (or
inference) are the low mass X-ray binary (LMXB) sources. These systems consist
of an accreting NS and a lighter donor companion. In a more ``explosive'' class
of binaries the NS accumulates the material until a critical density and
temperature are reached causing the ignition of the fuel (mostly H and He). The
fuel gets burned in a thermonuclear explosion within tens or hundreds of
seconds. The bursts are separated then by hours or tens of hours of quiescence.
From the measured energy and time structure of the associated X-ray bursts it is
possible to infer the mass and radius of the NS \cite{LAT07,Gal08,ozel16}.  This
inference proceeds, however, through some assumptions (e.g. about the
composition of the NS atmosphere) and needs additional information like the
temperature and distance to the source.

%
%
%
%
\begin{figure}[!htb]
\vspace*{-2mm}
 \begin{minipage}[h]{0.56\linewidth}
 \leavevmode
 \centering
  \includegraphics[width=\linewidth]{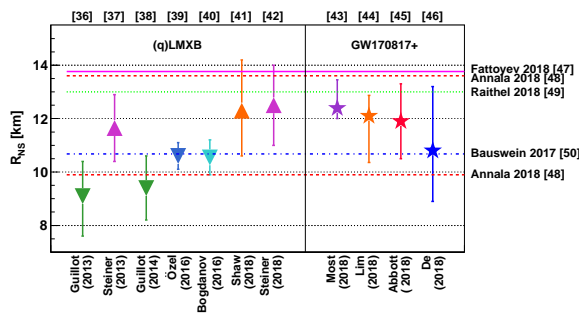}
 \end{minipage}
 \hfill
 \begin{minipage}[h]{0.43\linewidth}
 \leavevmode
 \centering

\caption{$R_{NS}$ from different analyses of (q)LMXB and GW170817 (triangles and
stars, respectively). The symbols correspond to refs.
\cite{gui13,ste13,gui14,ozel16,bog16,shaw18,ste18,mos18,lim18,abb18,de18} from
left to right. The lines represent the upper and lower limits from the early
analyses of GW170817. They correspond to refs. \cite{fat18,ann18,rai18,bau17} from top
to bottom.}
\label{fig_rns} 

 \end{minipage}
\vspace*{-2mm}

\end{figure}

Recent measurements of the mass and radius of NS come mainly from the analyses
of the thermal X-ray spectra from a more ``static'' class of LMXB, the
transiently accreting binaries in quiescence (qLMXB). First analyses of the same
five qLMXB from globular clusters gave however slightly inconsistent results for
the NS radii: $R_{NS} = 9.1^{+1.3}_{-1.5}$ km  \cite{gui13} and $10.4 <R_{NS} <
12.9$ km \cite{ste13} (the two leftmost points in Fig. \ref{fig_rns}). The 
differences in radii may be due to different assumptions about the composition
of the atmosphere, about the constancy of the radius for all NS, the distance
uncertainty and different statistical inference methods. This dichotomy seems to
persist up to now despite the subsequent reanalyzes including more qLMXB
sources: $R_{NS} = 9.4 \pm 1.2$ km \cite{gui14} or adding also thermonuclear
bursters: $10.1 <R_{NS} < 11.1$ km \cite{ozel16}, $9.9 <R_{NS} < 11.2$ km
\cite{bog16}, $R_{NS} = 12.3^{+1.9}_{-1.7}$ km  \cite{shaw18} and $11 < R_{NS} <
14$ km \cite{ste18}, see Fig. \ref{fig_rns}.




\section{Gravitational wave constraints}

The first gravitational wave signal from the neutron star merger event \cite{gw}
has triggered immediately a lot of interpretations. We will focus on a few of
them. The main constraint from the GW170817 event comes from the tidal
deformability, $\Lambda$, which is related to the strength of the quadrupole
mass deformation of a star due to the stress  caused by its companion's gravity
and obviously depends on the EoS. It was found to be $\Lambda < 800$ in
\cite{gw}. This constraint together with the requirement that the EoS should
support the 2 M$_{\odot}$ neutron stars yielded the following limits for the
radius of the 1.4 M$_{\odot}$ NS: $9.9 < R_{NS} < 13.6$ km in \cite{ann18}. It
very well matches the larger radius estimates from (q)LMXBs (see Fig.
\ref{fig_rns}). Ref. \cite{rai18} obtained the values of $\Lambda$ for a few EoS
supporting the $R_{NS}$ values from 10 to 15 km. From the Bayesian inference the
authors extracted the upper limit for the radius to be $R_{NS} < 13$ km and the
most likely one $R_{NS} \simeq 11.7$ km. GW170817 allowed also an estimate of
the lower limit for the NS radius from the fact that the merger did not result
in a prompt collapse, but instead some post merger electromagnetic emissions
have been observed. This lower limit has been estimated to be $R_{NS} > 10.68$
km in \cite{bau17}. Extraction of tidal deformabilities for 10 relativistic mean
field EoS taking into account the constraints from GW170817 allowed to estimate
the upper limit for the radius to be $R_{NS} < 13.76$ km in \cite{fat18}. Here
the $L$ value has also been specified for the corresponding TAMUC-FSUa EoS: $L <
82.5$ MeV. The upper and lower limits from the above analyses are represented in
Fig. \ref{fig_rns} by horizontal lines.

To complete the survey we quote the four most recent results constrained by the
GW170817 event. Using a numerous family of the EoS and the Bayesian inference
with the information on the lower and upper bounds on $\Lambda$ as well as on
the maximum mass of the NS, a most probable value of $R_{NS} =
12.39^{+1.06}_{-0.39}$ km has been obtained in \cite{mos18}. Similar analysis in
\cite{lim18} gave $R_{NS} = 12.10^{+0.77}_{-1.74}$ km. Refs. \cite{abb18} and
\cite{de18} attempted at deriving more tight constraints on $\Lambda$ than in
the original report \cite{gw} by applying an additional condition on the maximum
mass of the NS \cite{Dem10} and on the distribution of the measured
masses. They arrived at the values of $R_{NS} = 11.9^{+1.4}_{-1.4}$ km and
$R_{NS} = 10.8^{+2.4}_{-1.9}$ km, respectively. These results appear as stars in
Fig. \ref{fig_rns}. They reveal a better agreement with the larger radius
estimates from the binary X-ray systems.

\section{Squeezing the symmetry energy out of heavy ion collisions}

\begin{figure}[hbt] 
 \centering
\includegraphics*[width=.7\textwidth]{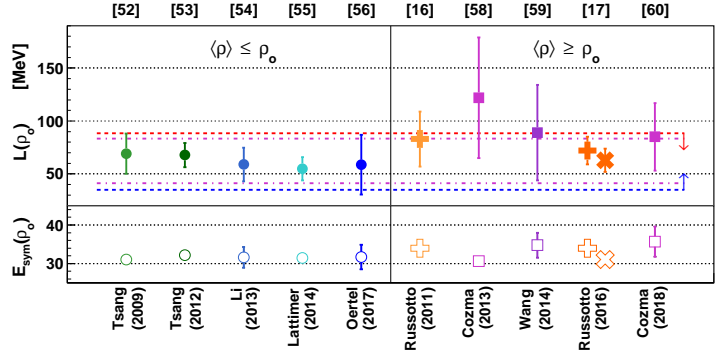}

\caption{Average values of the $E_{sym}(\rho_{o})$ (open symbols) and $L$
parameters (filled ones) obtained from the compilations
\cite{tsa09,tsa12,bali13,lat14,oert17} (circles). The cross, refs.
\cite{rus11,rus16}, and square, refs. \cite{coz13,wan14,coz18}, symbols
represent the results obtained from the analyses of the FOPI-LAND and ASY-EOS
data. The dashed lines represent the upper and lower limits for $L$ converted
from the $R_{NS}$ limits obtained for the GW170817 event in \cite{ann18} and
\cite{bau17}, respectively. The dash-dotted lines represent the 2$\sigma$
constraint from qLMXB\cite{ste13}.}

\label{fig_the} 
\end{figure}

The increasing amount of experimental data on the parameters of the
$E_{sym}(\rho)$ (53 have been collected in the most comprehensive review
\cite{oert17}) makes it more and more difficult to present them all together in
a clear way. Since 2009 there have been several attempts to do so. Here, instead
of quoting and discussing the individual results, we will show only the average
values from the selected reviews to trace the progress (see Fig. \ref{fig_the}).

The first compilation of the data available in 2009 has been presented in
\cite{tsa09}. It compares 7 results on the values of the $E_{sym}(\rho_{o})$ and
$L$ parameters obtained from the isospin diffusion, neutron and proton yields,
pygmy dipole resonances, PDR, and isobaric analogue states, IAS. The
corresponding points in Fig. \ref{fig_the} have been obtained in a similar way
as those in ref. \cite{bali13}, i.e. as simple mean values and average errors.
The extracted mean values for $E_{sym}(\rho_{o})$ and $L$ amount to
31.06$\pm$0.83 MeV and 69.16$\pm$19.06 MeV, respectively.

An update, including the results on binding energies, neutron skin thickness
from the elastic polarized proton scattering and on electric dipole
polarizabilities, EDP, as well as the results from the NS radius measurements was
presented in ref. \cite{tsa12}. The extracted mean values amount to
32.14$\pm$0.93 MeV and 67.9$\pm$11.4 MeV in this case.

The 28 results, including the analyses of atomic masses, isoscaling, IAS, skins,
PDR, isospin diffusion, transverse flow, EDP, NS data compiled in
\cite{bali13} yielded the average values of 31.6$\pm$2.66 and 58.9$\pm$16.0 MeV
for $E_{sym}(\rho_{o})$ and $L$, respectively.

The analysis of correlations between $L$ and $E_{sym}(\rho_{o})$
for 6 observables: binding energies, neutron skin thicknesses, dipole
polarizabilities, centroids of giant dipole resonances, isospin diffusion and
IAS resulted in the overlap values of $E_{sym}(\rho_{o})$ and $L$ of
31.45$\pm$1.05 MeV and 55$\pm$11 MeV, respectively \cite{lat14}.

Finally the most extensive analysis \cite{oert17} of 53 experimental results for
$E_{sym}(\rho_{o})$ and $L$ yielded the average values of 31.7$\pm$3.20 MeV and
58.7$\pm$28.1 MeV, respectively.

These results have been presented as circles in the left part of Fig.
\ref{fig_the}. All they have in common that they refer mostly to the observables
sensitive to densities around or below the saturation density. Moreover, these
experiments cannot claim to provide firm extrapolations to higher densities. 

The only way to study the properties of the asymmetric nuclear matter at high
densities in the laboratory conditions is to investigate the relativistic heavy
ion collisions. The only two available so far results coming from the high
energy experiments: FOPI-LAND \cite{rus11} and the ASY-EOS \cite{rus16}, aiming
at exploring the high density behavior of the $E_{sym}$ are depicted in the
right side of Fig. \ref{fig_the} as cross symbols. They were obtained as the
best fits of the model results to the experimental values of the elliptic flow
ratios for neutrons and hydrogens as a function of the transverse momentum. The
ratios have been obtained for the Au+Au collisions at 400 MeV/nucleon, at which
the squeeze-out of the matter out of the reaction plane attains a maximum. The
model that has been used here is the modified UrQMD version incorporating a
power-law-like density dependence of the $E_{sym}$ \cite{li05}. The values of
$L$ that have been obtained amount to 83$\pm$26 and 72$\pm$13 MeV for the
\cite{rus11} and \cite{rus16} analyses, respectively, both obtained for
$E_{sym}(\rho_{o})$=34 MeV. Lowering the $E_{sym}(\rho_{o})$ value to 31 MeV in
case of the ASY-EOS experiment resulted in a smaller value of $L$=63$\pm$11 MeV
(an X symbol in Fig. \ref{fig_the}). The figure shows also results obtained with
other models describing the experimental flow ratios. The T\"{u}QMD model gave
the value of $L=122\pm$57 MeV for a fixed $E_{sym}(\rho_{o})$=30.6 MeV
\cite{coz13} when applied to the FOPI-LAND data. Simulations performed with the
same model but with a modified momentum dependent potential when compared to the
ASY-EOS and FOPI-LAND results yielded a value of $L=85\pm$32 MeV for
$E_{sym}(\rho_{o})$ in the range $35.7\pm$3.9 MeV \cite{coz18}. An UrQMD model
with various Skyrme forces \cite{wan14} was able to best describe the FOPI-LAND
data with the $L=89\pm$45 MeV for  $E_{sym}(\rho_{o})$ extracted from the Skyrme
parameters in the range of 34.75$\pm$3.25 MeV.

The dashed horizontal lines in Fig. \ref{fig_the} represent the upper and lower
limits for $L$ obtained by converting the $R_{NS}$ limits from the GW170817
event in \cite{ann18} and \cite{bau17}, respectively. The approximate conversion
has been done using the phenomenological pressure-radius relation from Fig. 8 of
\cite{LAT16}. The obtained $L$ values for $R_{NS} = 13.6$ km \cite{ann18} and
$R_{NS} = 10.68$ km \cite{bau17} amount to 88.5 and 34.9 MeV, respectively. The
symmetry pressure, $p(\rho_{o}) = \rho_{o} L/3$. Finally, the dash-dotted lines
represent the 2$\sigma$ constraint on $L$ from the qLMXB analysis of
\cite{ste13} (see also Fig. \ref{fig_rns} for the corresponding $R_{NS}$).

The high density results are generally more stiff than the most recent averages
from the low energy experiments and from the nuclear data, which effectively
probe densities below $\rho_{o}$. Most of them comply with the GW170817
constraint (within the precision of the $R_{NS} \rightarrow L$ conversion
procedure) and also with the 2$\sigma$ constraint from qLMXB\cite{ste13}. On the
other hand, the values of the symmetry energy constant from the analyses of the
FOPI-LAND and the ASY-EOS data are generally higher than those from the
systematics. This might partially account for the observed differences in $L$
and be due to the correlation between the $L$ and $E_{sym}(\rho_{o})$ (see in
particular the two results for \cite{rus16}, where the result for a reduced
$E_{sym}(\rho_{o})$=31 MeV (the X symbol in Fig. \ref{fig_the}) gets much closer
to the recent averages for $L$). Nevertheless, since the high energy experiments
indeed have a chance to probe the densities above $\rho_{o}$, some differences
might be expected. 

The higher values of $L$ from the high energy experiments may support the
hypothesis on the soft to stiff transition at supra-normal densities
\cite{GAN12, GAN16}. Quantum Monte-Carlo simulations of \cite{GAN12} predict
that densities of up to 5$\rho_{o}$ can be reached in the centers of the NS.
Moreover, they predict that the uncertainty in the measured $E_{sym}$ of $\pm$2
MeV may lead to an uncertainty as large as 3 km for the radius of the NS. These
two predictions call for high precision measurements of the symmetry energy at
high densities (energies) to probe the EoS in the core of the NS and to provide
the results relevant for astrophysics.

\section{ASY-EOS II @ FAIR (202?)} 

A future project aiming at improving the ASY-EOS results has to take into
account two main goals: improvement of the precision, mainly through the usage of
high resolution detectors and radioactive beams, and extension of the density
region probed, through the usage of high energy beams. Microscopic simulations
of \cite{bali02} predict that the beams of energies around $\sim$1 GeV/nucleon
should be sufficient to attain the densities of up to $\sim$3$\rho_{o}$ in the
central zone of a heavy-ion collision. 

\begin{figure}[!htb]
 \begin{minipage}[h]{0.5\linewidth}
 \leavevmode
 \centering
  \includegraphics[width=\linewidth]{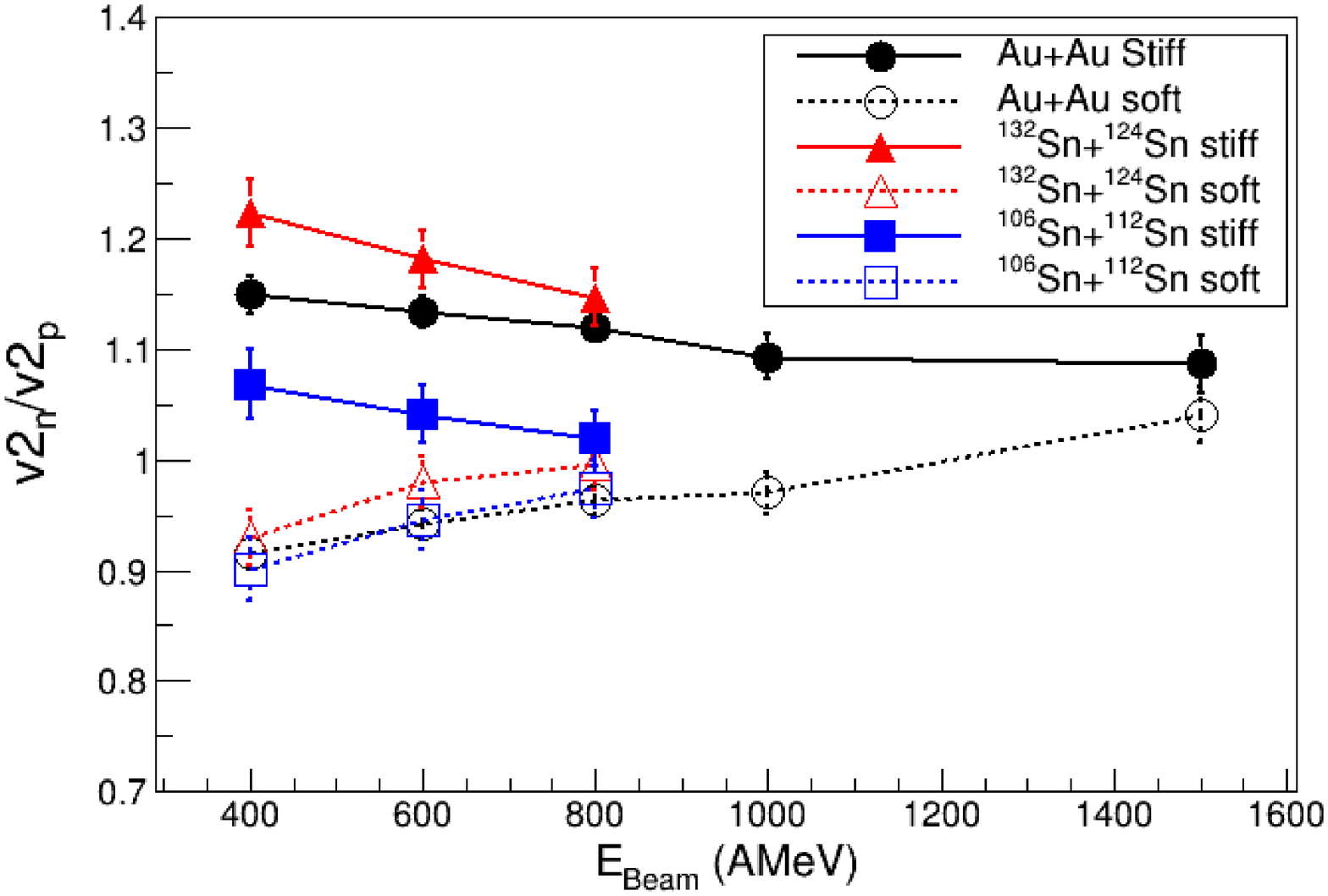}
 \end{minipage}
 \hfill
 \begin{minipage}[h]{0.5\linewidth}
 \leavevmode
 \centering
  \includegraphics[width=\linewidth]{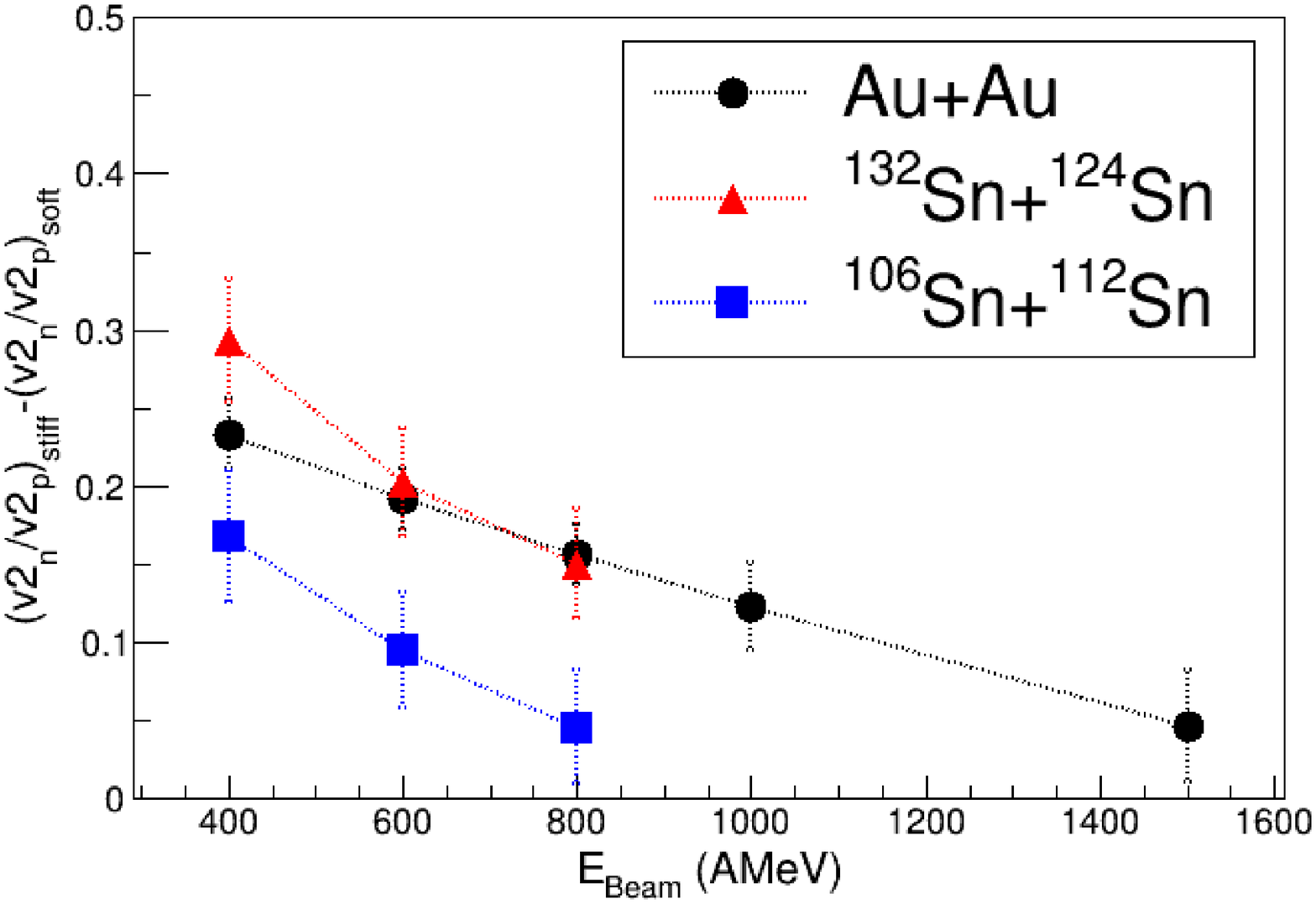}

 \end{minipage}
 
  \caption{ Left panel: Excitation functions of neutron-to-proton elliptic flow
  ratios, v2n/v2p as predicted by the UrQMD model for stiff and soft
  $E_{sym}(\rho)$. Right panel: differences between the stiff and soft results.
  Simulations and results by P. Russotto.} 

  \label{fig_pred}
\vspace*{-7mm}

\end{figure}

Simulations of semi-central Au+Au collisions at energies between 0.4 and 1.5
AGeV as well as neutron rich $^{132}$Sn+$^{124}$Sn and neutron poor
$^{106}$Sn+$^{112}$Sn systems at 0.4 - 0.8 AGeV have been carried out by using
the same version of the UrQMD transport code that has already been used in
\cite{rus16}. The neutron-to-proton elliptic flow ratio, v2n/v2p, at
mid-rapidity ($0.4 < y_{lab}/y_{proj} < 0.6$), with a stiff and a soft
parametrization of the potential part of the $E_{sym}$  for semi-central
($b_{red} < 0.54$) collisions are shown, as a function of the incident beam
energy, in the left panel of Fig. \ref{fig_pred}. The difference of such
v2n/v2p  ratios between the stiff and soft choices can be taken as a
sensitivity of the proposed observable, and is shown in the right panel of the
same figure.

The obtained sensitivity decreases with the beam energy, since the mean-field
contribution decreases at higher energies where the two-body collisions start to
dominate. Nevertheless, up to 1 AGeV the sensitivity of the proposed observable
is ~15\%, while a measurement can easily reach a ~5\% accuracy, allowing clear
discrimination between stiff and soft choices.
It is also important to stress the differences in trends (slopes) observed in
the left panel of Fig. \ref{fig_pred}, which by themselves have a discriminating
power. For the soft EoS the ratios increase with the energy while for the stiff
one the trend is opposite. This proves the needs for measuring the excitation
functions of these observables and the importance of using neutron rich beams
where the effect is stronger. In addition, measuring the double Sn+Sn system
would allow differential observables to be built which enable to control the
Coulomb vs $E_{sym}$ competition and to strongly reduce the model dependencies
and systematic errors. 

In order to improve the results as compared to the ASY-EOS ones it is necessary
to improve the mass resolution of the measured hydrogen isotopes. This seems to
be granted by the usage of the NeuLAND detector \cite{NEU11}, which would be the
main detector measuring neutrons in the first place. Additional charged particle
detectors like KRAB (see below), FOPI Plastic wall \cite{GOB93}, KRATTA \cite{LUK13}, FARCOS
\cite{PAG16} and CALIFA \cite{COR14} would allow to collect information on
centrality, reaction plane, light cluster production, flows and correlations.
The systems/energies intended to be measured in the future campaign are: 
$\bullet$ $^{197}$\textnormal{Au} + $^{197}$\textnormal{Au} at 400, 600, 1000 AMeV, 
$\bullet$ $^{132}$\textnormal{Sn} + $^{124}$\textnormal{Sn} at 400, 600  AMeV and
$\bullet$ $^{106}$\textnormal{Sn} + $^{112}$\textnormal{Sn} at 400, 600  AMeV.  
The setup requires a detector which would provide a fast trigger, based on the
multiplicity threshold and would precisely measure the azimuthal distributions
of charged particles beyond the angular acceptance of the FOPI Plastic Wall. A
schematic design of such a device is presented below.

\section{KRAB}

\begin{wrapfigure}{L}{0.5\textwidth}
 \centering
\captionsetup{width=.53\textwidth}
\includegraphics*[width=.53\textwidth]{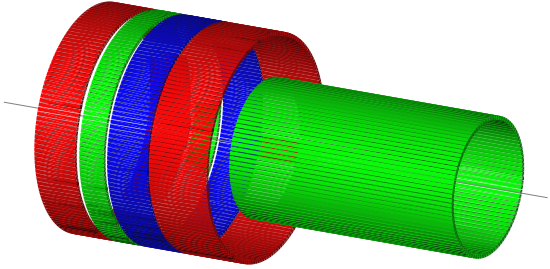}
\caption{Design of the KRAB detector.}
\label{fig_krab} 
\vspace*{-5mm}
\end{wrapfigure}

Taking into account the experimental demands, the current design of the
KRAk\'{o}w Barrel, KRAB, detector (see Fig. \ref{fig_krab}) assumes the following features: 
$\bullet$ 5 rings of 4$\times$4 mm$^{2}$ fast scintillating fibers (e.g. BCF-10) read out by SiPMs, 
$\bullet$ coverage of polar angles from 30$^{\circ}$ to 165$^{\circ}$,
$\bullet$ segmentation assuring more or less uniform count rates for the Au+Au at 1 AGeV, 
$\bullet$ geometrical efficiency $\sim$85\%
$\bullet$ less than 11\% of charged particles involved in multi-hits, 
$\bullet$ single segment multi-hit probability less than 5\%,
$\bullet$ sufficiently large entrance for radioactive beams,
$\bullet$ sufficiently small size and weight not to disturb neutrons,
$\bullet$ min radius $\sim$7 cm, 
$\bullet$ max radius $\sim$12 cm,
$\bullet$ length $\sim$46 cm,
$\bullet$ 4$\times$160 segments in forward rings,
$\bullet$ 96 segments in backward ring,
$\bullet$ 736 channels.

\noindent The above preliminary design and properties have been obtained using
the GEANT4 simulations with the UrQMD predictions for the Au+Au collisions at 1
AGeV used as an event generator. Based on the predicted multiplicity and angular
distributions it was possible to come out with the proposed structure and
segmentation. In particular, the simulations indicate that the Barrel should
provide better estimates of centrality as compared to the FOPI Plastic Wall. In
the first ASY-EOS experiment the target region was covered by the MICRO-BALL
\cite{SAR96} detector which was found very useful for providing a veto
information for reactions occurring on air and material up-stream of the target.
Nevertheless, it was too slow to be used as a trigger (CsI crystals) and had too
small segmentation for the multiplicities from the high energy beams. It was
also too sensitive to the high energy delta electrons, responsible for false
multiplicities.

Based on the experience from the first ASY-EOS campaign the design of the KRAB
detectors will try to overcome the observed problems and drawbacks. Thanks to
the high granularity, together with the FOPI Plastic Wall it should provide
sharp estimates for the orientation of the reaction plane and the unbiased
multiplicities. The observed quality of the simulated signals gives rise to the
expectation that the device will indeed provide very precise information on the
orientation of the reaction plane and on the centrality of the collision. It
will also play invaluable role by vetoing the upstream reactions.

The KRAB detector will cover about 85\% of the total solid angle. Together with
the FOPI Plastic Wall (~7\%) they will cover almost 92\% of the 4$\pi$.

Application of the fast plastic scintillator and fast silicon photon
counters together with the compact size of the KRAB detector will assure perfect
timing, and thus triggering properties and insensitivity to magnetic fields.
Compact size should allow to place the device inside bigger ones, such as e.g.
the CALIFA barrel.\\


Summarizing, we have shown the importance and difficulties of the symmetry
energy quest. Some controversies regarding the inference of the NS radius from
the qLMXB systems seem to be resolved by the recent interpretations of the first
gravitational wave signal. The results from low energy experiments and nuclear
data analyses on the stiffness of the symmetry energy ($L$ parameter) are well
within the limits imposed by the GW170817 signal and agree within 2$\sigma$ with
the results from the NS radius estimates. The values of $L$ parameter from high
energy measurements are generally slightly above the averages from the recent
compilations of the data from low density probes. Does it imply a transition to
more stiff EoS above the saturation density? Possibly the new planned
experiments will shed some more light on that. Definitely more tight constraints
on the symmetry energy, especially at high densities are needed. New missions,
new facilities, new experiments and new detectors herald new exciting results in
the nearest future.

\acknowledgments
\begin{center}
Work supported by Polish National Science Centre, \\
contract No. UMO-2017/25/B/ST2/02550
\end{center}


\end{document}